\documentclass[aip,pop,reprint,showpacs,superscriptaddress]{revtex4-1}
\usepackage{graphicx}
\usepackage{dcolumn}
\usepackage{bm}
\begin{document}


\title{Suppressing longitudinal double-layer oscillations by using elliptically polarized laser pulses
       in the hole-boring radiation pressure acceleration regime}
\author{Dong Wu}
\email{wudongphysics@gmail.com}
\affiliation{Center for Applied Physics and Technology,\\Peking University, Beijing, 100871, China.}
\affiliation{Key Laboratory of High Energy Density Physics Simulation, Ministry of Education,\\Peking University, Beijing, 100871, China.}
\author{C. Y. Zheng}
\email{zheng\_chunyang@iapcm.ac.cn}
\affiliation{Center for Applied Physics and Technology,\\Peking University, Beijing, 100871, China.}
\affiliation{Key Laboratory of High Energy Density Physics Simulation, Ministry of Education,\\Peking University, Beijing, 100871, China.}
\affiliation{Institute of Applied Physics and Computational Mathematics,\\Beijing, 100088, China.}
\author{C. T. Zhou}
\affiliation{Center for Applied Physics and Technology,\\Peking University, Beijing, 100871, China.}
\affiliation{Key Laboratory of High Energy Density Physics Simulation, Ministry of Education,\\Peking University, Beijing, 100871, China.}
\affiliation{Institute of Applied Physics and Computational Mathematics,\\Beijing, 100088, China.}
\author{X. Q. Yan}
\affiliation{Center for Applied Physics and Technology,\\Peking University, Beijing, 100871, China.}
\affiliation{Key Laboratory of High Energy Density Physics Simulation, Ministry of Education,\\Peking University, Beijing, 100871, China.}
\author{M. Y. Yu}
\affiliation{Institute of Fusion Theory and Simulation,\\Zhejiang University, Hangzhou, 310027, China.}
\author{X. T. He}
\email{xthe@iapcm.ac.cn}
\affiliation{Center for Applied Physics and Technology,\\Peking University, Beijing, 100871, China.}
\affiliation{Key Laboratory of High Energy Density Physics Simulation, Ministry of Education,\\Peking University, Beijing, 100871, China.}
\affiliation{Institute of Applied Physics and Computational Mathematics,\\Beijing, 100088, China.}
\date{\today}
\begin{abstract}
It is shown that well collimated mono-energetic ion beams with a large particle number can be generated in the hole-boring radiation pressure acceleration regime by using an elliptically polarized laser pulse with appropriate theoretically determined laser polarization ratio. 
Due to the $\bm{J}\times\bm{B}$ effect, 
the double-layer charge separation region is imbued with hot electrons that prevent
ion pileup, thus suppressing the double-layer oscillations.
The proposed mechanism is well confirmed by Particle-in-Cell simulations, and after suppressing the longitudinal double-layer oscillations, 
the ion beams driven by the elliptically polarized lasers own much better energy spectrum than those by circularly polarized lasers.

\end{abstract}
\maketitle

Well-collimated mono-energetic ion beams are useful for producing high
energy density matter, radiographing
transient processes, tumor therapy, and
ion-fast ignition in laser fusion 
\cite{PhysRevLett.106.145002, PhysRevLett.100.225001, PhysRevLett.86.436, PhysRevLett.85.2945, PhysRevLett.89.175003, PhysRevLett.88.215006}.
Radiation pressure acceleration (RPA) is an efficient scheme for generating high
quality ion beams, which as a particular case of laser-induced cavity pressure acceleration (LICPA)
\cite{PhyPla.19.053105,Appl.Phys.Lett.96.251502,Appl.Phys.Lett.99.071502} usually call for relatively high laser intensity. According to the target thickness, usually there are two modes of RPA acceleration mechanisms: 
light sail (LS) RPA for thin target 
\cite{PhysRevLett.103.024801, PhysRevLett.102.145002, PhysRevLett.105.155002, PhysRevLett.105.065002, PhysRevLett.105.065002,
PhysRevLett.100.135003, PhysRevLett.103.135001, PhyPla.16.044501, PhysRevLett.108.225002}
and hole boring (HB) RPA for thick target
\cite{PhyPla.18.056701, PhysRevLett.102.025002, PhyPla.16.083130, PlaPhyConFus.51.024004, PlaPhyConFus.51.095006, PhyPla.14.073101, PhyPla.18.073101, PhyPla.16.033102, PhysRevLett.106.014801}. 
In particular, the HBRPA owns the intrinsic property for large particle number acceleration \cite{PhyPla.18.053108}.
In HBRPA, the ponderomotive force of a laser pulse drives most or all
the local electrons inward, resulting in a shock-like
double-layer (DL) region with huge electrostatic charge-separation field. The latter traps and reflects the ions ahead of the DL, and compresses and accelerates
them inward like a piston \cite{PhyPla.18.056701, PhysRevLett.102.025002, PhyPla.16.083130, PlaPhyConFus.51.024004, PlaPhyConFus.51.095006, PhyPla.14.073101, PhyPla.18.073101, PhyPla.16.033102}. For RPA usually a circularly polarized (CP) laser
pulse is invoked \cite{PhyPla.18.056701, PhysRevLett.102.025002, PhyPla.16.083130, PlaPhyConFus.51.024004, PlaPhyConFus.51.095006, PhyPla.14.073101, PhyPla.18.073101, PhyPla.16.033102, PhysRevLett.106.014801}, since its ponderomotive force does not have a
high-frequency oscillating component that can preheat the electrons
and thus reduce the electrostatic DL field and efficiency of ion
acceleration. However, as pointed out by previous works \cite{PhyPla.16.033102, PlaPhyConFus.51.024004, PhyPla.16.083130}, the intense electrostatic DL field oscillates because trapping and reflection of ions by the this field are accompanied by consecutive increase and decrease of the ion therein. The oscillations lead to significant increase of the energy spread of the accelerated ions. On the other hand, if a linearly polarized (LP) laser pulse is used, the oscillating electrostatic DL field will be smeared out by the hot electrons generated by the second-harmonic component of the ponderomotive force. However, the DL structure is also destroyed and no mono-energetic ion beam is generated \cite{PhyPla.16.083130}.

In this paper, we propose to use an elliptically polarized (EP)
laser pulse for ion acceleration in the HBRPA regime. In this scheme,
$\bm{J}\times\bm{B}$ electron heating causes the DL region to be
filled with an appropriate distribution of hot electrons such that large ion density variations and electrostatic field oscillations in the DL region are suppressed, so that a mono-energetic ion beam can be generated.
An analytical model is used to estimate the most suitable laser polarization ratio
needed for suppressing the oscillating electrostatic field without destroying the DL structure.
The proposed mechanism is well confirmed by Particle-in-Cell (PIC) simulations, and the carbon ion beams driven by the elliptically polarized lasers own much better energy spectrum than those by circularly polarized lasers.

\begin{figure}
\includegraphics[width=8.50cm]{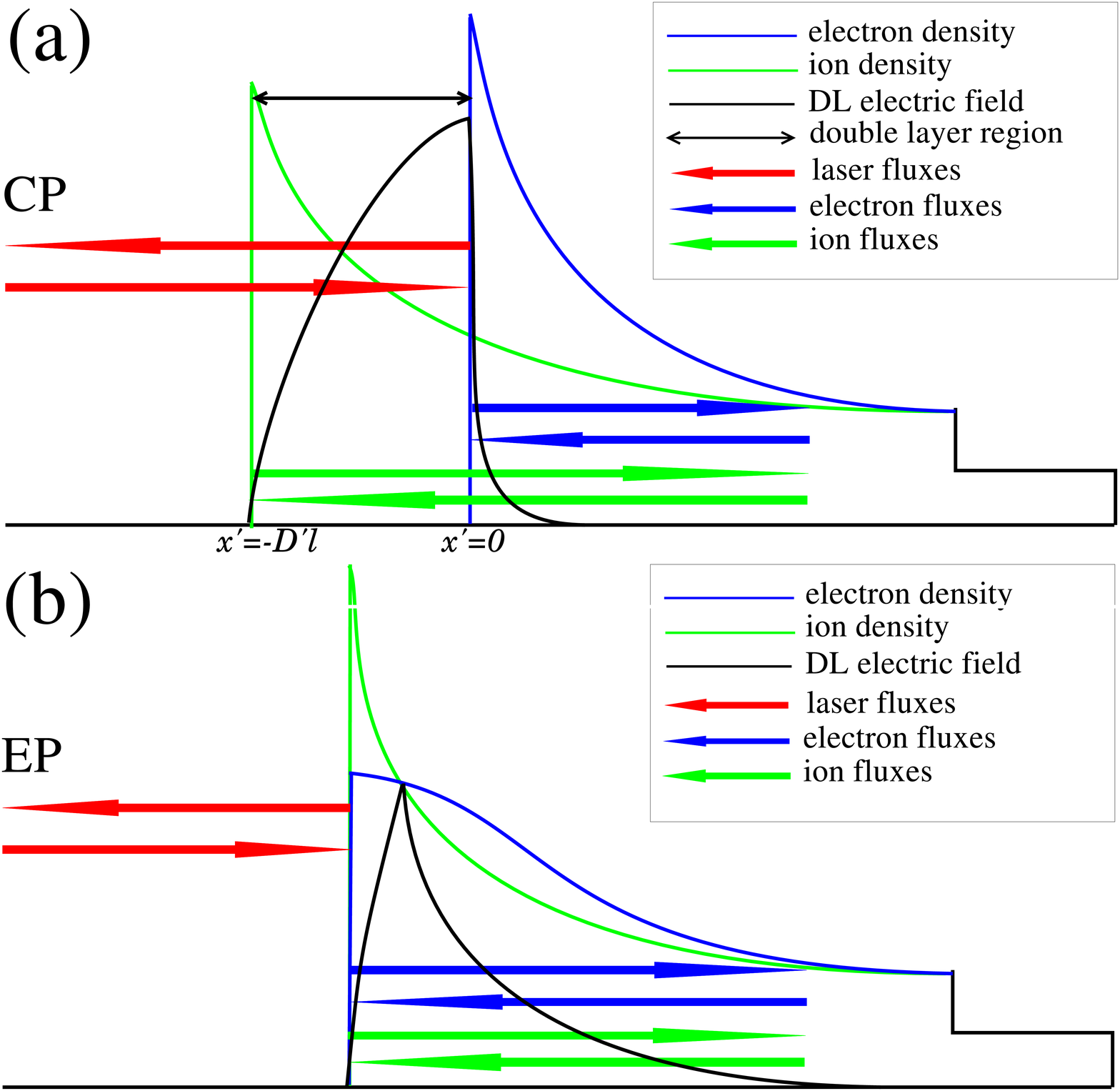}
\caption{\label{f1} (color online) Schematic structure of the electrostatic DL maintained by the radiation pressure 
in the piston-rest frame, for (a) CP laser and (b) EP laser.}
\end{figure}

\begin{figure}
\includegraphics[width=8.5cm]{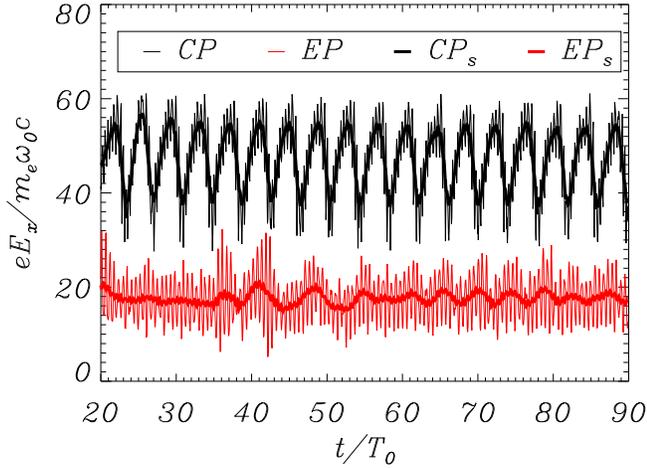}
\caption{\label{f2} (color online) The maximum electrostatic DL field $E_x$ ($x^{\prime}=0$) vs. time,
where the thin black (red) curve is for CP (EP) laser pulses. The thick black (red) line shows
the maximum electrostatic DL field in the absence of the high-frequency components of the CP (EP) lasers. The laser intensity is $4.4\times10^{21}$ W/cm$^{2}$ with $1.0$ $\mu$m wavelength, carbon plasma density is $n_e=20n_c$ and the ion charge number is $Z=6$. For EP laser pulse, the polarization ratio is $\alpha=0.64$.}
\end{figure}

\begin{figure}
\includegraphics[width=8.5cm]{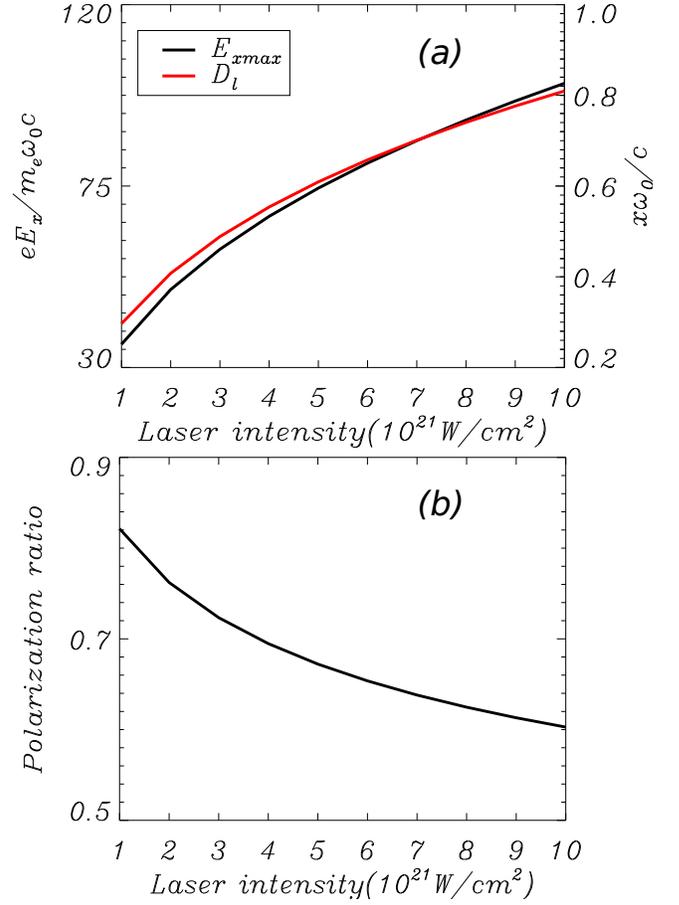}
\caption{\label{f3} (color online) (a) The maximum electrostatic DL field and width according to Eq.\ (6) and (7) vs. laser intensity, here the DL width has already transformed to the one in laboratory frame. (b) The optimum polarization
ratio $\alpha=a_z/a_y$ vs. the laser intensity.
The carbon plasma density $n_e=20n_c$ and ion charge number $Z=6$ is given.}
\end{figure}

Fig.\ 1 (a) illustrates the classical HBRPA scheme \cite{PhysRevLett.102.025002, PhyPla.16.083130}. 
The ponderomotive force of a laser pulse drives the local electrons inward, resulting in a shock-like
DL region with huge electrostatic charge-separation field. This electrostatic field traps and reflects the ions ahead of the DL, and compresses and accelerates them inward like a piston. 
We consider in the piston-rest frame \cite{PhysRevLett.102.025002, PhyPla.16.083130}. The ions which incident toward the DL with the velocity $\beta_f$ are reflected at $x^{\prime}=-D^{\prime}_l$,
where $\beta_f$ is the piston moving velocity and $D^{\prime}_l$ is the width of the DL in the piston-rest frame.
The electrons and laser light are reflected at the $x^{\prime}=0$.
From the momentum balance we have \cite{PhysRevLett.102.025002, PhysRevLett.102.025002}
\begin{equation}\label{1}
I_{0}(1-\beta_f)/(1+\beta_f)=n_i(m_i+Zm_e)c^3\gamma_f^2\beta_f^2,
\end{equation}
where $m_i$, $m_e$, $n_i$, and $Z$ are the ion mass, electron mass, ion density, and ion charge number, respectively, $\gamma_f=1/\sqrt{1-\beta_f^2}$,
$I_0=a_0^2m_en_cc^{3}$ is the CP laser intensity, and $a_0=eE_0/m_e\omega_0c$ is the normalized CP laser amplitude. The velocity of the piston normalized by the light speed $c$ can then be written as
$\beta_f=\sqrt{I_{0}/n_i(m_i+Zm_e)c^{3}}/$ $[1+\sqrt{I_{0}/n_i(m_i+Zm_e)c^{3}}]$.
As the ion acceleration takes place mainly in the DL region, the equation of motion of an ion particle within the DL is
\cite{PhyPla.16.083130}
\begin{equation}\label{2}
m_ic^2d\gamma^{\prime}_i/dx^{\prime}=ZeE_x,
\end{equation}
where $\gamma^{\prime}_i=1/\sqrt{1-\beta^{'2}_i}$ and $E_x$ is the electrostatic DL field. 
The Poisson equation is
\begin{equation}\label{3}
dE_x/dx^{\prime}=4{\pi}Zen^{\prime}_i.
\end{equation}
Ion continuity leads to
$n^{\prime}_i=2n_i\gamma_f\beta_f/\beta^{\prime}_i$. From Eqs.\ (2)
and (3), we can then obtain
\begin{equation}\label{4}
d^2\gamma^{\prime}_i/dx^{\prime2}=
2\omega^2_{pi}\gamma_f\beta_f/c^2\beta^{\prime}_i.
\end{equation}
At the reflection point $x^{\prime}=-D^{\prime}_l$, we have
$\beta^{\prime}_i=0$, so that $\gamma^{\prime}_i=1$. Charge conservation leads to $E_x(x=-D^{\prime}_l)=0$ or $d\gamma^{\prime}_i/d{x^{\prime}}=0$.

The first integral of Eq.\ (4) after multiplying $d\gamma^{\prime}_i/d{x^{\prime}}$ on both sides is
\begin{equation}\label{5}
d\gamma^{\prime}_i/dx^{\prime}=2\omega_{pi}\sqrt{\gamma_f\beta_f}(\gamma^{\prime2}_i-1)^{1/4}/c,
\end{equation}
which relates the ion kinetic energy to the electrostatic DL field. 
At the position $x^{\prime}=0$ of peak electron density,
$\gamma^{\prime}_i$ can be approximated by
$\gamma^{\prime}_i(x^{\prime}=0)=\gamma_f$ since $E_x$ outside the DL
region is very small and has little effect on the incident velocity
$\beta_f$. Accordingly, the maximum electrostatic DL field $E_x$ (at $x^{\prime}=0$) is
\begin{equation}\label{6}
E_{x\max}=(m_ic^2/Ze)d\gamma^{\prime}_i/dx^{\prime}=2m_ic\omega_{pi}\beta_f\gamma_f/Ze.
\end{equation}
Integrating Eq.\ (5) and assuming $\beta_f\ll 1$, we can obtain the
double layer width approximately as \cite{PhyPla.16.083130}
\begin{equation}\label{7}
D^{\prime}_l=\beta_{f}c/3\omega_{pi}\gamma_f,
\end{equation}
which corresponds to $D_l=\gamma_fD^{\prime}_l=\beta_{f}c/3\omega_{pi}$ in the laboratory frame.

The analysis above assumes quasistationarity in the piston-rest
frame. In reality, as can be seen in Fig.\ 2, the maximum electrostatic DL
field $E_x$ is oscillating \cite{PhyPla.16.033102, PlaPhyConFus.51.024004, PhyPla.16.083130}. In
fact, in the piston-rest frame $E_x$ does not instantaneously reflect
the ions. Instead, ion acceleration involves a finite time \cite{PlaPhyConFus.51.024004, PhyPla.16.083130}.
When the ions stream into the DL region, the total charge density grows and a
large electrostatic field is induced. However, when the ions are reflected, the total charge density and the electrostatic field
decrease. Thus the sawtooth structure of the electrostatic field
is closely associated with the steep variation of the ion density in the
DL region. The large oscillating electrostatic DL field disperses the accelerated
ions \cite{PlaPhyConFus.51.024004}, which would otherwise be mono-energetic. 

One can expect that if the reflections of ions and electrons are synchronized, as depicted in Fig.\ 1 (b), the sharp variations of the ion density and oscillations of the electrostatic DL field can be reduced or eliminated. Here we propose to accomplish this by using an EP laser pulse. In this case, the laser intensity in Eq.\ (1) is replaced by $I_0=(a^2_y+a^2_z)m_en_cc^{3}/2$. The $\bm{J}\times\bm{B}$ effect of the EP laser pulse drives the electrons out of the $x^{\prime}=0$ plane, and fills the double layer region with hot electrons. The longitudinal electron momentum is
\cite{splpi} $p_{e}=(a^2_y-a^2_z)/4$, with $a_y>a_z$. The piston structure can be maintained if these electrons are stopped (in the forward direction) within the DL. Balancing the (forward) kinetic and the electrostatic and ponderomotive potential energies of the electrons, we can obtain the most suitable polarization ratio $\alpha$. The ponderomotive force acting on the electrons is \cite{PlaPhyConFus.51.024014}
\begin{equation}\label{8}
2(I_{0}/c)(1-\beta_f)/(1+\beta_f)=\int{f_pn^{\prime}_e(x^{\prime})dx^{\prime}},
\end{equation}
where the integral can be approximated by $\bar{f_p}n_ec/\omega_{pe}$. Thus, the averaged ponderomotive force $\bar{f_p}$ is
\begin{equation}\label{9}
\bar{f_p}=2\omega_{pe}(I_{0}/n_e c^2)(1-\beta_f)/(1+\beta_f).
\end{equation}
From the energy balance relation, we can get the optimum coupling condition
\begin{equation}\label{10}
\int\limits_{-D^{\prime}_l}^{0}eE_x(x^{\prime})dx^{\prime}+m_ec^2(\gamma_{e}-1)
=\bar{f_p}D^{\prime}_l,
\end{equation}
where $\gamma_{e}=\sqrt{1+p^2_{e}}$. According to Eq.\ (2), the integral can be approximated by $m_ic^2(\gamma_f-1)/Z$ if the effect of the very weak $E_x$ outside the DL region is neglected. The condition Eq.\ (10) can then be written as
\begin{widetext}
\begin{eqnarray}\label{11}
m(\gamma_f-1)/Z+\sqrt{1+(a^2_y-a^2_z)^2/16}-1=\sqrt{m/Z}(a^2_y+a^2_z)\beta_f(1-\beta_f)/3(1+\beta_f)\gamma_fn,
\end{eqnarray}
\end{widetext}
where $m=m_i/m_e$ and $n=n_e/n_c$.
For given plasma density and laser intensity which is $I_0=(a^2_y+a^2_z)m_en_cc^{3}/2$, the optimum polarization ratio $\alpha$ (which equals $a_z/a_y$) for the proposed scheme can then be obtained from Eq.\ (11). Fig.\ 3 (b) shows the optimum polarization ratio $\alpha$ as a function of the laser intensity for given carbon plasmas with density $n_e=20n_c$ and ion charge number $Z=6$.

We have applied a series of one dimensional (1D) PIC (\texttt{KLAP} code \cite{Eur.Phys.Lett.95.55005, PhysRevLett.107.265002}) simulations to further confirm this mechanism. As for the 
longitudinal problem, 1D PIC is enough to demonstrate the main physics. The size of the simulation is $80\lambda_0$ with $\lambda_0$ represents the laser wavelength. The simulation box is divided into uniform grid of $16000$. 
The CP laser pulse propagates into the simulation box from the left boundary. The thick target consists of two species: electrons and carbon ions with ion charge number $z=6$, which are initially located in the region $10\lambda_0 < x < 40\lambda_0$ with density
$n_e = 20n_c$, where $n_c=\omega_0^2e^2m_e/4\pi=1.1\times10^{21}$/cm$^3$ is the critical density for $1.0$ $\mu$m wavelength laser pulses. We use $600$ electrons and $100$ carbon ions per cell to run the simulations. 
The normalized amplitude of the CP laser electric field is $a_y=a_z=40$, corresponding to the laser intensity $4.4\times10^{21}$W/cm$^2$. The CP laser pulse is of constant temporal intensity profile. In contrast, the EP laser pulses with the
same intensity and temporal profile are also run. Here in our simulations, we have scan the EP polarization ratio $\alpha$ to obtain the most suitable one with the narrowest energy spectrum, which is $\alpha=0.64$. While according to Eq.\ (11), for the given laser intensity and carbon ion density, the theoretical predicted one is $\alpha=0.68$, which is quite close to the PIC optimized one.

\begin{figure}
\includegraphics[width=8.5cm]{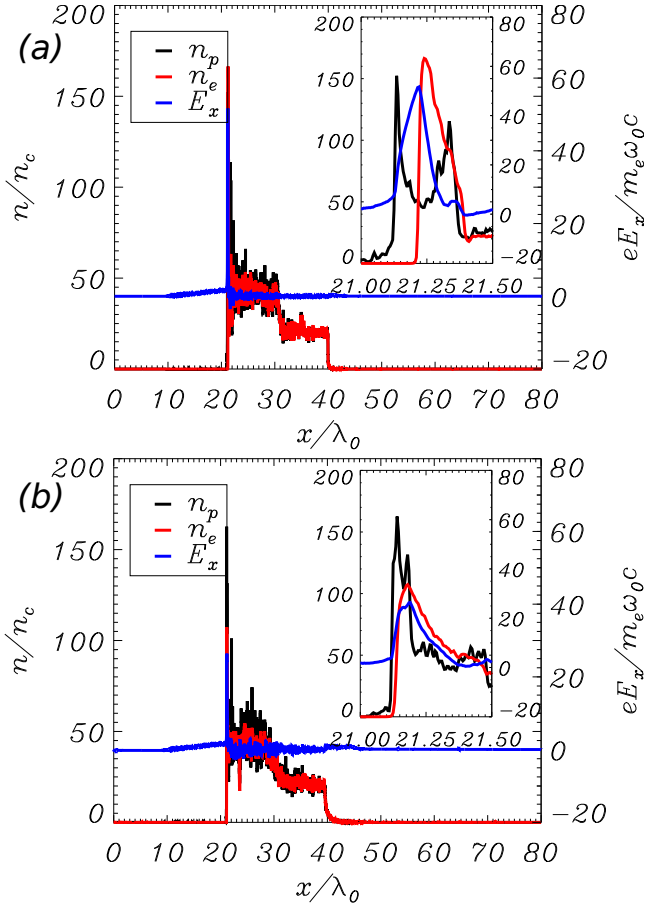}
\caption{\label{f4} (color online) The electron, carbon ion density and electrostatic DL field profiles corresponding to the
CP and EP laser pulses, respectively, at $t=50T_0$. The simulation parameters are the same as for Fig.\ 2. The insets are magnified views of the DL regions.}
\end{figure}

\begin{figure}
\includegraphics[width=8.5cm]{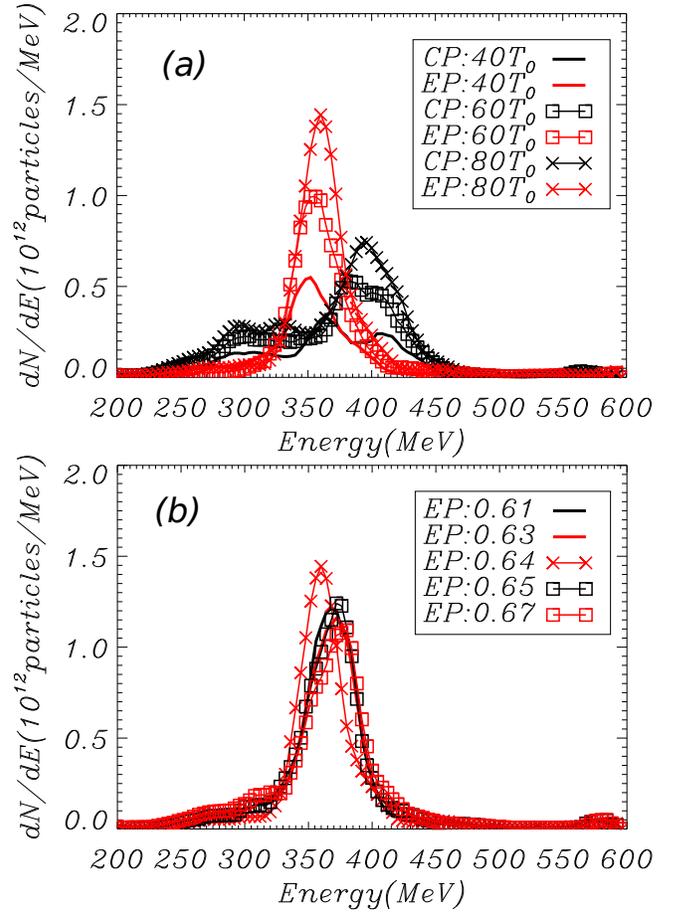}
\caption{\label{f5} (color online) Energy spectrum the carbon ion beams from PIC simulations.
(a) The energy spectrum of proton beams driven by CP and EP laser pulses at $t=40T_0$, $t=60T_0$ and $t=80T_0$, in which the polarization ratio is $\alpha=0.64$.
(b) The energy spectrum of the carbon ion beams driven by EP laser pulses with polarization ratio $\alpha=$ $0.61$, $0.63$, $0.64$, $0.65$, $0.67$ at $t=80T_0$.
Here, the laser intensity is $4.4\times10^{21}$ W/cm$^{2}$ with $1.0$ $\mu$m wavelength, carbon plasma density is $n_e=20n_c$ and the ion charge number is $Z=6$.}
\end{figure}
Fig.\ 4 (a) and (b) show the profiles of the electron, carbon ion densities and the electrostatic DL field driven by the CP and EP laser pulses, respectively. We see that these profiles are consistent with the physical pictures as shown in Fig.\ 1. 
The effective ion-acceleration electrostatic field corresponding to the CP laser pulse is mostly within the DL region. From our simulation the maximum electrostatic DL field and width for CP laser pulse are about $60.0$ normalized unit and $0.10\lambda_0$ which agree well with the theoretical predicted ones as shown in Fig.\ 3 (a). As discussed, the electrostatic DL field is sawtooth oscillating. On the other hand, when the EP laser pulse is used, the reflection of electrons and ions take place together. In this case the width of the DL becomes practically zero. However, the space-charge electric field is nonzero since the electron density at the reflection point (region) remains lower than the ion density. In fact, it has a long tail in the laser propagation direction as the ions and electrons move together. We can see in Fig.\ 2 that the ion-accelerating electrostatic DL field is indeed not oscillating, which is the key to realizing the mono-energetic ion beams.

Here we go to the details of the oscillating electrostatic DL field. Fig.\ 2 (c) shows the time evolution of the maximum DL electric field for both the CP (black curve) and EP (red curve) laser pulses.
Two oscillation components can be identified. The high-frequency component is induced by the electron motion, and the low-frequency
oscillations at several laser periods are induced by the trapping and
reflection of the ions \cite{PlaPhyConFus.51.024004}. Because of their large mass,
the carbon ions cannot affect nor respond to the high-frequency component of
the electric field. That is, the ions respond only to the time averaged (over the fast variations) electric field, as shown by the thick black and red curves. We can see that the sawtooth electric field profile corresponding to the CP laser remains, but that corresponding to the EP laser is greatly suppressed.

As expected, after suppressing the longitudinal electrostatic DL field using EP laser pulses, the carbon ion beam quality is greatly improved compared with that of CP laser pulses, which is clearly shown in Fig.\ 5 (a). For the ion beams driven by EP laser pulses, the peak energy can be as high as $360$ MeV with half-width-half-maximum energy spread less than $5.0\%$. 
To test the robust of this scheme, EP laser pulses with polarization ratio $0.67>\alpha>0.61$ have been run, the quality of the carbon ion beams remain much better than that driven by the CP laser pulse with the same intensity and temporal profile, which is shown in Fig\ 5 (b).  

To achieve the fast ignition (FI) driven by ion beams, the specification of the
beam quality is as follows \cite{PhyPla.16.102701, PlaPhyConFus.51.035010, PlaPhyConFus.51.014008}: $10^{16}$ proton particles with energy $3\sim5$ MeV or carbon
ion $10^{14}$ with energy $300\sim500$ MeV. The relatively smaller particle number makes the FI driven by carbon ions more attractable, but it calls for higher standard of the energy spread. For example, if the peak energy of carbon
ion beams is $400$ MeV and the distance (from ion source to the compressed core) is $2$ cm, $10\%$ and $50\%$
energy spreads correspond to $12.5$ kJ and great than $20$ kJ ignition energy \cite{PhyPla.16.102701}. We believe that the EP laser driven HBRPA is a potential way to realize the specification of the beam quality for FI, which is of large particle number and narrow energy spread.

It should be mentioned that the scheme has also been applied to proton-acceleration, and the EP laser pulses are confirmed to be quite efficient than the CP laser pulses. 
If the laser wavelength is longer, say $10.6$ $\mu$m of CO$_2$ lasers, the laser intensity can be reduced to below $10^{20}$ W/cm$^2$, but the plasma density should be reduced accordingly \cite{PhysRevLett.106.014801, Nat.Phys.8.95}. From the relations (1) and (11), the quality of the ion beams would not change, but the particle number would be significantly less.

For thin target, usually the LS model of RPA dominates. 
However, the acceleration mechanisms of LSRPA and the HBRPA are quite different. For LSRPA, the electrons and ions form a neutral layer, which, as a whole, is continually accelerated by the light pressure. Once the neutrality of the layer is ruined, the ion particles would undergo Coulomb explorations, and the acceleration would terminate. The CP laser, with constant ponderomotive force, can well keep the neutrality of the layer, which means that the CP laser is the best choice for LSRPA \cite{New.J.Phys.10.113005}. 
The energy and intensity of accelerated ion beams diminish 
with decreasing $\alpha$ and almost disappear for $\alpha<0.7$.

In summary, we have proposed to use an EP
laser pulse for ion acceleration in the HBRPA regime. An analytical model is used to estimate the most suitable laser polarization ratio. In this scheme,
$\bm{J}\times\bm{B}$ electron heating causes the DL region to be
filled with an appropriate distribution of hot electrons such that large ion density variations and electrostatic field oscillations in the DL region are suppressed, so that a mono-energetic ion beam can be generated.
This scheme is further confirmed by 1D PIC simulations, and the ion beams driven by the EP lasers own much better energy spectrum than those by CP lasers.


\begin{acknowledgments}
One of the authors Dong Wu thanks B. Qiao and F. L. Zheng for fruitful discussions.
This work was supported by the National Natural Science Foundation of
China (Grant Nos. 11075025, 10835003, 10905004, 11025523, and 10935002) and the Ministry of Science and Technology of China (Grant No. 2011GB105000).
\end{acknowledgments}


\begin{thebibliography}{99}

\bibitem{PhysRevLett.106.145002} C. Wang, X.-T. He, and P. Zhang, Phys. Rev. Lett. 106, 145002 (2011).

\bibitem{PhysRevLett.100.225001} C. K. Li, F. H. S ́guin, J. R. Rygg, J. A. Frenje, M. Manuel, R. D. Petrasso, R. Betti, 
J. Deletetrez, J. P. Knauer, F. Marshall, D. D. Meyerhofer, D. Shvarts, V. A. Smalyuk, C. Stoeckl, 
O. L. Landen, R. P. J. Town, C. A. Back, and J. D. Kilkenny, Phys. Rev. Lett. 100, 225001 (2008).

\bibitem{PhysRevLett.86.436} M. Roth, T. E. Cowan, M. H. Key, S. P. Hatchett, C. Brown, W. Fountain, J. Johnson, D. M.
Pennington, R. A. Snavely, S. C. Wilks, K. Yasuike, H. Ruhl, F. Pegoraro, S. V. Bulanov,
E. M. Campbell, M. D. Perry, and H. Powell, Phys. Rev. Lett. 86, 436 (2001).

\bibitem{PhysRevLett.85.2945} R. A. Snavely, M. H. Key, S. P. Hatchett, T. E. Cowan, M. Roth, T. W. Phillips, M. A. Stoyer,
E. A. Henry, T. C. Sangster, M. S. Singh, S. C. Wilks, A. MacKinnon, A. Offenberger, D. M. Pennington, K. Yasuike, A. B. Langdon,
B. F. Lasinski, J. Johnson, M. D. Perry, and E. M. Campbell, Phys. Rev. Lett. 85, 2945 (2000).

\bibitem{PhysRevLett.89.175003} T. Z. Esirkepov, S. V. Bulanov, K. Nishihara, T. Tajima, F. Pegoraro, V. S. Khoroshkov,
K. Mima, H. Daido, Y. Kato, Y. Kitagawa, K. Nagai, and S. Sakabe, Phys. Rev. Lett. 89, 175003 (2002).

\bibitem{PhysRevLett.88.215006} A. J. Mackinnon, Y. Sentoku, P. K. Patel, D. W. Price, S. Hatchett, M. H. Key, C. Andersen,
R. Snavely, and R. R. Freeman, Phys. Rev. Lett. 88, 215006 (2002).

\bibitem{PhyPla.19.053105} J. Badziak, S. Jablonski, T. Pisarczyk, P. Raczka, E. Krousky, R. Liska, M. Kucharik, 
T. Chodukowski, Z. Kalinowska, P. Parys, M. Rosinski, S. Borodziuk, and J. Ullschmied,
Phys. Plasmas 19, 053105 (2012).

\bibitem{Appl.Phys.Lett.96.251502} J. Badziak, S. Borodziuk, T. Pisarczyk, T, Chodukowski, E. Krousky, K.
Masek, J. Skala, J. Ullschmied, and Y.-J. Rhee, Appl. Phys. Lett. 96, 251502 (2010).

\bibitem{Appl.Phys.Lett.99.071502} J. Badziak and S. Jabłonski, Appl. Phys. Lett. 99, 071502 (2011).

\bibitem{PhysRevLett.103.024801} M. Chen, A. Pukhov, T. P. Yu, and Z. M. Sheng, Phys. Rev. Lett. 103, 024801 (2009).

\bibitem{PhysRevLett.102.145002} B. Qiao, M. Zepf, M. Borghesi, and M. Geissler, Phys. Rev. Lett. 102, 145002 (2009).

\bibitem{PhysRevLett.105.155002} B. Qiao, M. Zepf, M. Borghesi, B. Dromey, M. Geissler, A. Karmakar, and P. Gibbon, 
Phys. Rev. Lett. 105, 155002 (2010).

\bibitem{PhysRevLett.105.065002} T.-P. Yu, A. Pukhov, G. Shvets, and M. Chen, 
Phys. Rev. Lett. 105, 065002 (2010).

\bibitem{PhysRevLett.100.135003} X. Q. Yan, C. Lin, Z. M. Sheng, Z. Y. Guo, B. C. Liu, Y. R. Lu, J. X. Fang, and J. E. Chen,
Phys. Rev. Lett. 100, 135003 (2008).

\bibitem{PhysRevLett.103.135001} X. Q. Yan, H. C. Wu, Z. M. Sheng, J. E. Chen, and J. Meyer-ter Vehn, Phys. Rev. Lett. 103,
135001 (2009).

\bibitem{PhyPla.16.044501} X. Q. Yan, M. Chen, Z. M. Sheng, and J. E. Chen, 
Phys. Plasmas 16, 044501 (2009).

\bibitem{PhysRevLett.108.225002} C. A. J. Palmer, J. Schreiber, S. R. Nagel, N. P. Dover, C. Bellei, F. N. Beg, S. Bott, 
R. J. Clarke, A. E. Dangor, S. M. Hassan, P. Hilz, D. Jung, S. Kneip, S. P. D. Mangles, K. L. Lancaster, 
A. Rehman, A. P. L. Robinson, C. Spindloe, J. Szerypo, M. Tatarakis, M. Yeung, M. Zepf, and Z. Najmudin1, 
Phys. Rev. Lett. 108, 225002 (2012).

\bibitem{PhyPla.18.056701} A. P. L. Robinson, Phys. Plasmas 18, 056701 (2011).

\bibitem{PhysRevLett.102.025002} N. Naumova, T. Schlegel, V. T. Tikhonchuk, C. Labaune, I. V. Sokolov, and G. Mourou,
Phys. Rev. Lett. 102, 025002 (2009).

\bibitem{PhyPla.16.083130} T. Schlegel, N. Naumova, V. T. Tikhonchuk, C. Labaune, S. I. V., and G. Mourou, 
Phys. Plasmas 16, 083103 (2009).

\bibitem{PlaPhyConFus.51.024004} A. P. L. Robinson, P. Gibbon, M. Zepf, S. Kar, R. G. Evans, and C. Bellei, 
Plasma Phys. and Controlled Fusion 51, 024004 (2009).

\bibitem{PlaPhyConFus.51.095006} A. P. L. Robinson, D.-H. Kwon, and K. Lancaster, 
Plasma Phys. and Controlled Fusion 51, 095006 (2009).

\bibitem{PhyPla.14.073101} X. M. Zhang, B. F. Shen, X. M. Li, Z. Y. Jin, and F. C. Wang, Phys. Plasmas 14, 073101 (2007).

\bibitem{PhyPla.18.073101} Xiaomei Zhang, Baifei Shen, Liangliang Ji, Wenpeng Wang, Jiancai Xu, Yahong Yu, and Xiaofeng Wang,
Phys. Plasmas 18, 073101 (2011).

\bibitem{PhyPla.16.033102} Xiaomei Zhang, Baifei Shen, Zhangying Jin, Fengchao Wang, and Liangliang Ji,
Phys. Plasmas 16, 033102 (2009).

\bibitem{PhysRevLett.106.014801} Charlotte A. J. Palmer, N. P. Dover, I. Pogorelsky, M. Babzien, G. I. Dudnikova, M. Ispiriyan, M. N. Polyanskiy, J. Schreiber, P. Shkolnikov, V. Yakimenko, and Z. Najmudin,
Phys. Rev. Lett. 106, 014801 (2011).

\bibitem{PhyPla.18.053108} J. Badziak, G. Mishra, N. K. Gupta, and A. R. Holkundkar,
Phys. Plasmas 18, 053108 (2011).

\bibitem{splpi} P. Gibbon, Short Pulse Laser Interactions with Matter (Imperial College, London, 2005).

\bibitem{PlaPhyConFus.51.024014} V K Tripathi , C S Liu , X Shao , B Eliasson and R Z Sagdeev,
Plasma Phys. and Controlled Fusion 51, 024014 (2009).

\bibitem{Eur.Phys.Lett.95.55005} F. L. Zheng, S. Z. Wu, C. T. Zhou, H. Y. Wang, X. Q. Yan, and X. T. He,
Eur. Phys. Lett. 95, 55005 (2011).

\bibitem{PhysRevLett.107.265002} H. Y. Wang, C. Lin, Z. M. Sheng, B. Liu, S. Zhao, Z. Y. Guo, 
Y. R. Lu, X. T. He, J. E. Chen and X. Q. Yan, 
Phys. Rev. Lett. 107, 265002 (2011).

\bibitem{PhyPla.16.102701} J. J. Honrubia, J. C. Fernández, M. Temporal, B. M. Hegelich, and J. Meyer-ter-Vehn,
Phys. Plasmas 16, 102701 (2009). 

\bibitem{PlaPhyConFus.51.035010} M. Temporal, R. Ramis, J. J. Honrubia, and S. Atzeni, 
Plasma Phys. Controlled Fusion 51, 035010 (2009).

\bibitem{PlaPhyConFus.51.014008} J. J. Honrubia and J. Meyer-ter-Vehn, 
Plasma Phys. Controlled Fusion 51, 014008, (2009).

\bibitem{Nat.Phys.8.95}Dan Haberberger, Sergei Tochitsky, Frederico Fiuza, Chao Gong, Ricardo A. Fonseca,
Luis O. Silva, Warren B. Mori and Chan Joshi, Nat. Phys. 8, 95, (2011). 

\bibitem{New.J.Phys.10.113005} S G Rykovanov, J Schreiber, J Meyer-ter-Vehn, C Bellei, A Henig, H C Wu and M Geissler,
New Journal of Physics 10, 113005 (2008). 

\end{thebibliography}
{}
\end{document}